\documentclass[11pt]{article}

\usepackage[top=50mm, bottom=50mm, left=50mm, right=50mm]{geometry}
\usepackage{amssymb}
\usepackage{amsmath}
\usepackage{amsthm}
\usepackage{amsfonts}
\usepackage{mathtools}
\usepackage{graphicx}
\usepackage{graphics}
\usepackage{float}
\usepackage{color}
\usepackage{fullpage}
\usepackage[normalem]{ulem} 
\usepackage{makeidx}
\usepackage{xspace}
\usepackage{authblk}
\usepackage{datetime}
\usepackage{microtype}
\usepackage{xpatch}
\usepackage{xstring}
\usepackage{fancyhdr}
\usepackage{mathrsfs}
\usepackage{esint}
\usepackage{upgreek}
\usepackage{siunitx}
\usepackage{chemformula}
\usepackage{tabularx, multirow} 
\usepackage{threeparttable}
\usepackage[]{booktabs} 
\usepackage{float}
\usepackage{graphicx,xcolor}
\usepackage[colorlinks=false]{hyperref}
\usepackage[noabbrev]{cleveref}
\usepackage[font = footnotesize, labelfont=bf]{caption}

\fancypagestyle{fpg}{%
	\lfoot{\footnotesize Energies \textbf{2019}, 12(22), 4360}%
	\cfoot{\normalsize\thepage}
	\rfoot{\footnotesize \href{doi.org/10.3390/en12224360}{DOI: 10.3390/en12224360}}%
}

\makeindex

\begin{document}

\title{Sensitivity of Numerical Predictions to the~Permeability~Coefficient in Simulations of Melting and Solidification Using the~Enthalpy-Porosity~Method}
 
	\author[1,$\dagger$]{Amin Ebrahimi}
	\author[2]{Chris R. Kleijn}
	\author{Ian M. Richardson}
	\affil[1]{Department of Materials Science and Engineering, Delft University of Technology, Mekelweg~2, 2628CD~Delft, The~Netherlands}
	\affil[2]{Department of Chemical Engineering, Delft University of Technology, van~der~Maasweg~9, 2629HZ Delft, The~Netherlands}
	\affil[$\dagger$]{Corresponding author, Email: A.Ebrahimi@tudelft.nl}
	
\date{}
\maketitle
\thispagestyle{fpg}

\begin{abstract}
		The high degree of uncertainty and conflicting literature data on the value of the permeability coefficient (also known as the mushy zone constant), which aims to dampen fluid velocities in the mushy zone and suppress them in solid regions, is a critical drawback when using the fixed-grid enthalpy-porosity technique for modelling non-isothermal phase-change processes. In the present study, the sensitivity of numerical predictions to the value of this coefficient was scrutinised. Using finite-volume based numerical simulations of isothermal and non-isothermal melting and solidification problems, the causes of increased sensitivity were identified. It was found that depending on the mushy-zone thickness and the velocity field, the solid--liquid interface morphology and the rate of phase-change are sensitive to the permeability coefficient. It is demonstrated that numerical predictions of an isothermal phase-change problem are independent of the permeability coefficient for sufficiently fine meshes. It is also shown that sensitivity to the choice of permeability coefficient can be assessed by means of an appropriately defined P{\'e}clet number.
\end{abstract}

\newpage

\section{Introduction}
\label{sec: introduction}

\noindent
Numerical simulations of melting and solidification processes are critical to develop our understanding of phase transformations that occur in various technologies such as additive manufacturing, thermal energy storage, anti-icing and materials processing. It is however challenging due to the involvement of the moving boundary problem~\cite{Crank1984} and the wide range of length and time scales in transport phenomena during solid--liquid phase transformations~\cite{Shyy_2002}. Different numerical techniques have been developed to resolve the transport phenomena and the solid--liquid phase transition at different scales, which have been reviewed in~\cite{Rappaz_1989,Dutil_2011,Verma_2014,Jaafar_2017}.

Techniques for numerically modelling solid--liquid phase transitions at the continuum level have generally been divided into transformed-grid and fixed-grid approaches~\cite{Dutil_2011}. Detailed information on the derivation and implementation of these approaches can be found in the literature~\cite{Basu_1988,Lacroix_1990,Voller_2009}. The~focus of the present work is on the fixed-grid approach in which latent heat effects and fluid flow near the liquid--solid interface are taken into account through the inclusion of thermal energy and momentum source terms in that region~\cite{Rappaz_1989}. One of the advantages of the fixed-grid approach over the transformed-grid approach is its robustness in treating changes in the interface topology. However, interface smearing is an inherent disadvantage, which may be diminished by applying local grid refinement near the solid--liquid interface~\cite{Mencinger_2004,Hannoun_2005,Lan_2002}.

Flow velocities in the liquid phase, particularly in the so-called mushy zone close to the solid interface, where phase-change takes place over a melting temperature range~\cite{Worster_1997}, can have a significant influence on the local heat transfer~\cite{Le_Quere_1999}; it is therefore crucial to predict fluid flow in the mushy zone with a sufficient level of accuracy. Of~the various approaches that have been employed for this purpose~\cite{Morgan_1981,Voller_1987,Voller_1987_2}, the~porosity approach is the most common. In~this approach, it is assumed that the fluid flow in the mushy zone is analogous to that in a porous medium. Consequently, Darcy's law is assumed to govern the  flow  and a corresponding sink term ($\vec{S}_\mathrm{d}$) is added to the momentum equation,
\begin{equation}
\vec{S}_\mathrm{d} = \frac{- \mu}{K} \vec{V},
\label{eq: damp_term}
\end{equation}

\noindent
where $\vec{V}$ is the fluid velocity vector and $\mu$ dynamic viscosity. Assuming isotropy for the solid--liquid morphology and using the Blake--Kozeny equation~\cite{Brinkman_1949}, the~permeability ($K$) can be defined as a~function of the liquid fraction ($f_\mathrm{L}$) as follows:
\begin{equation}
K = \frac{\mu}{C} \frac{f_\mathrm{L} ^3}{\left(1-f_\mathrm{L}\right)^2},
\label{eq: permeability}
\end{equation}

\noindent
with $C$ the so-called permeability coefficient (or the mushy zone constant). The~Blake--Kozeny equation is valid for liquid fractions lower than 0.7~\cite{Poirier_1987,Singh_2001}; however, it is typically used for the~entire range of liquid fractions. The~value of the permeability coefficient $C$ and its influence on numerical simulations of melting and solidification is generally~uncertain.	

The values reported for the permeability coefficient $C$ in the literature generally range between $10^3$ and $10^{15}\,\SI{}{\kilogram\per\second\per\meter\cubed}$ \cite{Fadl_2019,Hong_2019}; however, values between $10^4$ and $10^{8}\,\SI{}{\kilogram\per\second\per\meter\cubed}$ are often applied. {The~proposed values for $C$ do not seem to have a one-to-one relation to the type of material being studied. For~instance, values between $\mathcal{O}(10^4)$ \cite{Rai_2007} and $\mathcal{O}(10^8)$ \cite{Zheng_2014} have been proposed for stainless steel, and~values between $\mathcal{O}(10^4)$ \cite{Yang_2016} and $\mathcal{O}(10^{15})$ \cite{Kousksou_2014} have been proposed for gallium. \mbox{Thus, the~values} proposed for $C$ in the literature seem to depend markedly on the process parameters and the associated boundary conditions, and~thus lack generality}. In~other words, tuning the value of $C$ is essential for every set of boundary conditions and material properties, which requires considerable trial and error evaluation. Previous studies on the influence of the permeability coefficient in simulations of melting and solidification have mainly concentrated on finding a value for $C$ to diminish discrepancies between numerical and experimental data for a specific problem \mbox{(see, for~instance,}~\cite{Karami_2019,Pan_2018,Arena_2017,Prieto_2016,Kheirabadi_2015,Hosseinizadeh_2013,Shmueli_2010}). Additionally, they have often focused on the phase-change materials for thermal energy storage applications for which material properties and operating conditions differ from those for materials processing applications (\textit{e.g.},   casting, welding and additive manufacturing). It is necessary to know under which circumstance and to what extent the numerical predictions of melting and solidification are sensitive to the value of the permeability coefficient to~avoid the need for excessive computation. However, there is as yet no general guideline for assessing the appropriate value of $C$.

The degree of sensitivity of the numerical predictions to the value of the permeability coefficient $C$ appears to be diverse. Several studies reported that the chosen value of $C$ affect the numerical results predicted by the enthalpy-porosity method for isothermal phase-change problems~\cite{Kumar_2017,Sattari_2017,Hu_2014}, in~which phase transformation occurs at the melting temperature of the material. This effect is not physically realistic for isothermal phase transformations and should be considered as a numerical artefact~\cite{Vogel_2016}. Pan \textit{et al.}~\cite{Pan_2018} studied the melting of calcium chloride hexahydrate (\ch{CaCl2H12O6}) in a~vertical cylinder and stated that the value of $C$ depends on the temperature difference that drives melting and should be tuned to obtain a reasonable agreement with experimental data. Fadl and Eames~\cite{Fadl_2019} reported that the numerical predictions are less sensitive to the value of $C$ in regions where heat transfer is dominated by conduction. Assessing the degree of sensitivity of the numerical prediction to the value of $C$ for different materials and boundary conditions is an open question, despite its significance has been emphasised in the literature~\cite{Kheirabadi_2015,Hameter_2016,Hong_2019,Fadl_2019}.

Realising the degree of sensitivity of the numerical predictions to the value of the permeability coefficient in phase-change simulations with a particular interest in melting and solidification during welding and additive manufacturing  was the motivation for the present study. Our aim was to quantitatively assess the influence of the permeability coefficient ($C$) on the results of the enthalpy-porosity method in predicting heat and fluid flow and the position of the solid--liquid interface during solid--liquid phase transformations. The~sensitivity of the results to the permeability coefficient was analysed for both isothermal and non-isothermal melting and solidification problems, and~possible roots of errors in the simulation of solidification and melting processes were highlighted. Our study quantified the influence of the permeability coefficient on the numerical predictions as a function of numerical simulation parameters and physical process parameters and elucidated the limitations of the enthalpy-porosity~method.

\section{Problem~Description}
\label{sec: problemDescription}

\begin{figure}[H]
	\centering
	\includegraphics[height=0.2\textheight]{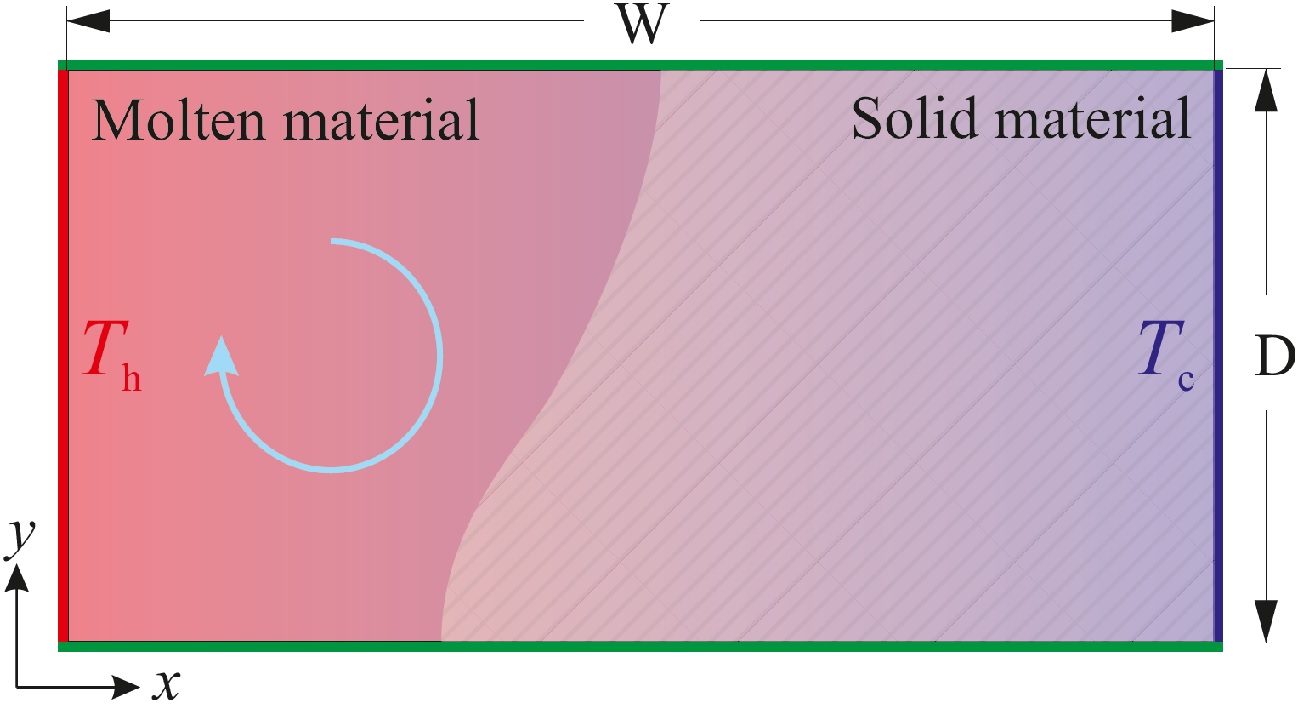}
	\caption{The schematic of the system under consideration. From $t = 0$, the~left vertical wall is kept at an isothermal temperature $T_\mathrm{h}$, and~the right vertical wall is kept at an isothermal temperature $T_\mathrm{c}$, whereas the horizontal walls are adiabatic.}
	\label{fig: schematic}
\end{figure}

\noindent
Isothermal and non-isothermal phase transformations in the  two-dimensional rectangular enclosure shown in \cref{fig: schematic} were studied. The~length of the enclosure (W) is twice its height ($\mathrm{D} = 0.1~\si{\meter}$). The~enclosure is initially filled with solid material at a temperature ($T_\mathrm{i}$) below the melting temperature $T_\mathrm{m}$ (for isothermal phase-change) or solidus temperature $T_\mathrm{s}$ (for non-isothermal phase-change). The~left and the right solid walls are isothermal walls and the upper and the lower walls are adiabatic. At~the starting time of the simulation, the~left wall temperature is suddenly raised from the initial temperature ($T_\mathrm{i}$) to the hot wall temperature ($T_\mathrm{h}$), whereas the cold wall is kept at a temperature $T_\mathrm{c}$ = $T_\mathrm{i}$. The~thermophysical properties of our artificial materials are presented in \cref{tab: material_properties}, which represent a wide range of metallic and non-metallic phase-change materials. The~liquid phase is assumed to be incompressible and Newtonian, with~constant dynamic viscosity $\mu$. All other thermophysical material properties are assumed to be the same for both the solid and liquid phases and temperature~independent.

\unskip

\begin{table}[H]
	\centering
	\caption{Thermophysical properties of the artificial materials used in the~simulations.}
	\begin{tabular}{lll}
		\toprule
		\bf{Property                                                                                   } & \bf{Value}               & \bf{Unit}                                \\ \midrule
		Density $\rho$                                                                              & $10^3$              & \si{\kilogram\per\meter\cubed}      \\
		Specific heat capacity $c_\mathrm{p}$                                                       & $10^3$              & \si{\joule\per\kilogram\per\kelvin} \\
		Thermal conductivity $k$                                                                    & 1, 10 and 100                & \si{\watt\per\meter\kelvin}         \\
		Dynamic viscosity $\mu$                                                                     & $10^{-3}$           & \si{\kilogram\per\meter\per\second} \\
		Latent heat of fusion $L_\mathrm{m}$                                                        & $10^5$              & \si{\joule\per\kilogram}            \\
		Thermal expansion coefficient  $\beta$                                                      & $10^{-6}$           & \si{\per\kelvin}                    \\
		Melting temperature $T_\mathrm{m} = \frac{\left(T_{\mathrm{l}} + T_{\mathrm{s}}\right)}{2}$ & \SI{1.5e3}{}        & \si{\kelvin}                        \\
		Melting-temperature range $\Delta T_{\mathrm{m}}$                                           & 0, 5, 10, 25 and 50 & \si{\kelvin}                        \\ \bottomrule
	\end{tabular} 
	\label{tab: material_properties}
\end{table}
\unskip

\section{Mathematical~Formulation}
\label{sec: mathematicalModel}

\noindent
Since the Rayleigh number for the problems considered here is less than $10^6$, the~fluid flow is assumed to be laminar. Thermal buoyancy effects are taken into account using the Boussinesq assumption~\cite{Tritton_1977}. Utilising the dimensionless variables $\mathrm{Fo} = t\alpha / \mathrm{D}^2$, $\vec{V}^* = \vec{V}\,\mathrm{D} / \alpha$, $T^* = T / \left(T_\mathrm{h} - T_\mathrm{melt}\right)$, \mbox{$H^* = T^* + f_\mathrm{L}/\mathrm{Ste}$}, $p^* = p\,\mathrm{D}^2 / \left(\rho\alpha^2\right)$ and $K^* = K / \mathrm{D}^2$ \cite{Mencinger_2004}, the~unsteady momentum and energy equations are cast in conservative dimensionless form, respectively, as~follows:
\begin{equation}
\frac{1}{\mathrm{Pr}} \left(\frac{\partial \vec{V}^*}{\partial \mathrm{Fo}} + \vec{V}^* \cdot \nabla\vec{V}^* + \nabla p^*\right) = 	
\nabla^2 \vec{V}^* + \mathrm{Ra}T^*\hat{y} - \frac{\vec{V}^*}{K^*},
\label{eq: momentum}
\end{equation}
\begin{equation}
\frac{\partial H^*}{\partial \mathrm{Fo}} + \vec{V}^* \cdot \nabla H^* = 
\nabla^2 T^* + \frac{1}{\mathrm{Ste}} \left(\frac{\partial f_\mathrm{L}}{\partial \mathrm{Fo}} + \nabla \vec{V}^*\right),
\label{eq: energy}
\end{equation}

\noindent
where the Prandtl number (Pr) represents the ratio of momentum diffusivity to thermal diffusivity $\left(\alpha = k / \left(\rho c_\mathrm{p}\right)\right)$ in molten materials and is defined as
\begin{equation}
\mathrm{Pr} = \frac{\mu}{\rho\alpha}.
\label{eq: prandtl}
\end{equation}

To evaluate the relative importance of buoyancy to viscous forces acting on the molten materials, the~Grashof number (Gr) and Rayleigh number (Ra) can be defined as
\begin{equation}
\mathrm{Gr} = \frac{g\beta \rho^2 \,\mathrm{D}^3 \left(T_\mathrm{h} - T_\mathrm{melt}\right)}{\mu^2},
\label{eq: grashof}
\end{equation}
\begin{equation}
\mathrm{Ra} = \mathrm{Gr}\cdot\mathrm{Pr} = \frac{g\beta \rho \,\mathrm{D}^3 \left(T_\mathrm{h} - T_\mathrm{melt}\right)}{\mu \alpha}.
\label{eq: rayleigh}
\end{equation}

The Stefan number (Ste) is the ratio of sensible to latent heat and is defined as
\begin{equation}
\mathrm{Ste} = \frac{c_\mathrm{p} \left(T_\mathrm{h} - T_\mathrm{melt}\right)}{L_\mathrm{m}}.
\label{eq: stefan}
\end{equation}

In the above, $\rho$ is density, $\mu$ dynamic viscosity, $p$ the static pressure, $k$ thermal conductivity, $c_\mathrm{p}$~specific heat capacity at constant pressure, $f_\mathrm{L}$ the local liquid fraction, $L_\mathrm{m}$ the latent heat of melting or solidification, $\beta$ the thermal expansion coefficient, $g$ the gravitational acceleration, $t$~time  and $\vec{V}$ the fluid velocity vector. $\hat{y}$ is the unit vector in the y-axis direction. $T_\mathrm{melt}$ is the~melting-temperature (for~isothermal phase-change) or solidus temperature (for non-isothermal phase-change).

For the sake of simplicity, the~local liquid fraction was considered to be a function of temperature only, which is a reasonable assumption for the cases where under-cooling is not significant~\cite{Voller_1990}. Different relationships have been proposed for temperature-dependence of the liquid fraction, depending on the materials and the nature of the micro-segregation~\cite{Voller_1991,Swaminathan_1992}. Based on the method suggested by Voller and Swaminathan~\cite{Voller_1991}, the~relationship between the local liquid fraction and the temperature is defined to be linear as follows:
\begin{equation}
f_\mathrm{L}\left(T\right) = \frac{T - T_\mathrm{s}}{T_\mathrm{l} - T_\mathrm{s}}; T_\mathrm{s} \le T \le T_\mathrm{l},
\label{eq: temp_dep_fL}
\end{equation}

\noindent
where  $T_\mathrm{l}$ and $T_\mathrm{s}$ are liquidus and solidus temperatures, respectively. In~the case of isothermal phase-transformation, a~step change in the liquid fraction occurs at the melting-temperature according to the method given by Voller and Prakash~\cite{Voller_1987}. Additionally, the~convective part of the enthalpy source term in the energy equation (\textit{i.e.},    $\nabla\vec{V}^*$) takes the value zero for the isothermal phase transformation due to the step change in the latent heat and a zero velocity at the solid--liquid interface~\cite{Voller_1987}. 

To deal with the fluid velocity in the mushy zone, a~sink term based on the Blake--Kozeny equation (\textit{i.e.}, \cref{eq: damp_term} and \cref{eq: permeability}) was introduced into the momentum equation~\cite{Voller_1987}.
\begin{equation}
\frac{- \vec{V}^*}{K^*} = - \frac{\mathrm{D}^2\,\vec{V}^*}{K} = - \frac{C}{\mu} \frac{\left(1-f_\mathrm{L}\right)^2}{f_\mathrm{L}^3+\epsilon} \,\mathrm{D}^2 \,\vec{V}^*,
\label{eq: sink_term_voller}
\end{equation}

\noindent
where $C$ is the permeability coefficient and $\epsilon$ is a small constant, here chosen to be equal to $10^{-3}$, to~avoid division by zero. The~sink term is zero in the liquid region $\left(f_\mathrm{L} = 1\right)$ while its limiting value for $f_\mathrm{L} = 0$ should be large enough to dominate the other terms in the momentum equation to~suppress the fluid velocities in the solid region. The~value of the permeability coefficient $C$ can be expected to depend on the morphology of the mushy zone. No consensus was found on the value of the permeability coefficient in the literature~\cite{Ferreira_2009,Faraji_2010,Bouabdallah_2012,Zheng_2014,Mahdaoui_2014,Farsani_2017}. The~effects of this parameter on numerical results are therefore reported~here.

The following dimensionless parameters are utilised to construct a framework for analysing our results. The~ratio $\theta$ of melting-temperature range $\left(\Delta T_\mathrm{m} = T_\mathrm{l} - T_\mathrm{s} \right)$ to the temperature difference between the hot and cold wall $\left(\Delta T_\mathrm{w} = T_\mathrm{h} - T_\mathrm{c} \right)$ is defined as
\begin{equation}
\theta = \frac{T_\mathrm{l} - T_\mathrm{s}}{T_\mathrm{h} - T_\mathrm{c}} = \frac{\Delta T_\mathrm{m}}{\Delta T_\mathrm{w}}.
\label{eq: theta}
\end{equation}

The volumetric fraction of liquid inside the computational domain at a time instant $t$ is evaluated~as
\begin{equation}
\phi (t) = \frac{1}{\mathrm{W}\cdot \mathrm{D}} \iint \limits_{\mathrm{domain}} f_\mathrm{L} (t) \mathop{}\!\mathrm{d}A.
\label{eq: phi}
\end{equation}

We   use the relative difference $\Delta\phi (t) = \left|\phi_2 (t) - \phi_1 (t) \right|/\phi_1 (t)$ between two solutions obtained using different permeability coefficients $C_2$ and $C_1$ to quantify the sensitivity of a solution to the chosen value of the permeability coefficient $C$. In~addition, a~dimensionless parameter ($\Gamma$) is employed to quantify the difference between the solid--liquid interface morphology predicted using different permeability coefficients $C_1$ and $C_2$ as follows:
\begin{equation}
\Gamma = \frac{1}{\mathrm{W}\cdot \mathrm{D}} \int \limits_{y = 0}^{y = D} \sqrt{\left(x_{\mathrm{f}, C_1} - x_{\mathrm{f}, C_2}\right)^2} \mathop{}\!\mathrm{d}y,
\label{eq: gamma}
\end{equation}

\noindent
where $x_{\mathrm{f}}$ is the position of the liquidus~line.

\section{Numerical~Procedure}
\label{sec: numericalProcedure}

\noindent
Numerical predictions obtained from an open-source (OpenFOAM) and a commercial (ANSYS Fluent) solver were compared for melting and solidification simulations, and~the results were found to be rather identical (see \cref{fig: validation}). ANSYS Fluent (Release 18.1) was selected to carry out the calculations, and~simulations were performed in parallel on eight cores of an Intel Xeon E5-2630 processor (2.20~GHz). The~computational domain was discretised on a uniform mesh with quadrilateral grid cells. After~performing a grid independence test (results presented in \cref{sec: gridSensitivityIsothermal,sec: gridSensitivityNonisothermal}), a~base grid with a cell size $\Delta x_\mathrm{i} / \mathrm{D} = \Delta y_\mathrm{i} / \mathrm{D} =  \SI{4e-3}{}$ was chosen. To~capture a sharp solid--liquid interface and to reduce errors associated with high-gradient regions, especially when the thickness of the mushy zone is smaller than the base grid size  (\textit{i.e.}, $\left(\Delta T_\mathrm{m} /\Delta T_\mathrm{w}\right) \mathrm{W} \le \Delta x_\mathrm{i}$ ), a~dynamic solution-adaptive mesh refinement was applied using the ``gradient approach''~\cite{Dannenhoffer_1985} based on the liquid fraction gradients. Four levels of mesh refinement $\left(\Delta x_\mathrm{N} = \Delta x_\mathrm{i} / 2^{\mathrm{N}}, \mathrm{N} = 1, \cdots, 4 \right)$ were applied every ten time-steps. A~fixed time-step size of $\Delta\mathrm{Fo} = k\Delta t / \left(\rho c_\mathrm{p} \Delta x_\mathrm{i} ^2\right) = 10^{-4}$ was selected, corresponding to  a Courant number $\left(\mathrm{Co} = |\vec{V}|\Delta t / \Delta x \right)$ less than~0.25.

The conservation equations were discretised using the finite-volume method. The~central- differencing scheme was utilised for the discretisation of the convection and diffusion terms both with second-order accuracy. The~Pressure-Implicit with Splitting of Operators (PISO) scheme~\cite{Issa_1986} was used for pressure--velocity coupling. Additionally, the~PRESTO (PREssure STaggering Option) scheme~\cite{Patankar_1980} was used for the pressure interpolation. The~time derivative was discretised with a second-order implicit scheme. Convergence requires that scaled residuals of the continuity, momentum and the energy equations fall below $10^{-10}$, $10^{-12}$ and $10^{-14}$, respectively, and~that the relative change in the volumetric fraction of liquid $\phi$ from one iteration to the next is less than $10^{-10}$.

\section{Results}
\label{sec: results}
\vspace{-6pt}

\subsection{Model~Verification}
\label{sec: modelVerification}

\begin{figure}[H]
	\centering
	\includegraphics[width=1.00\linewidth]{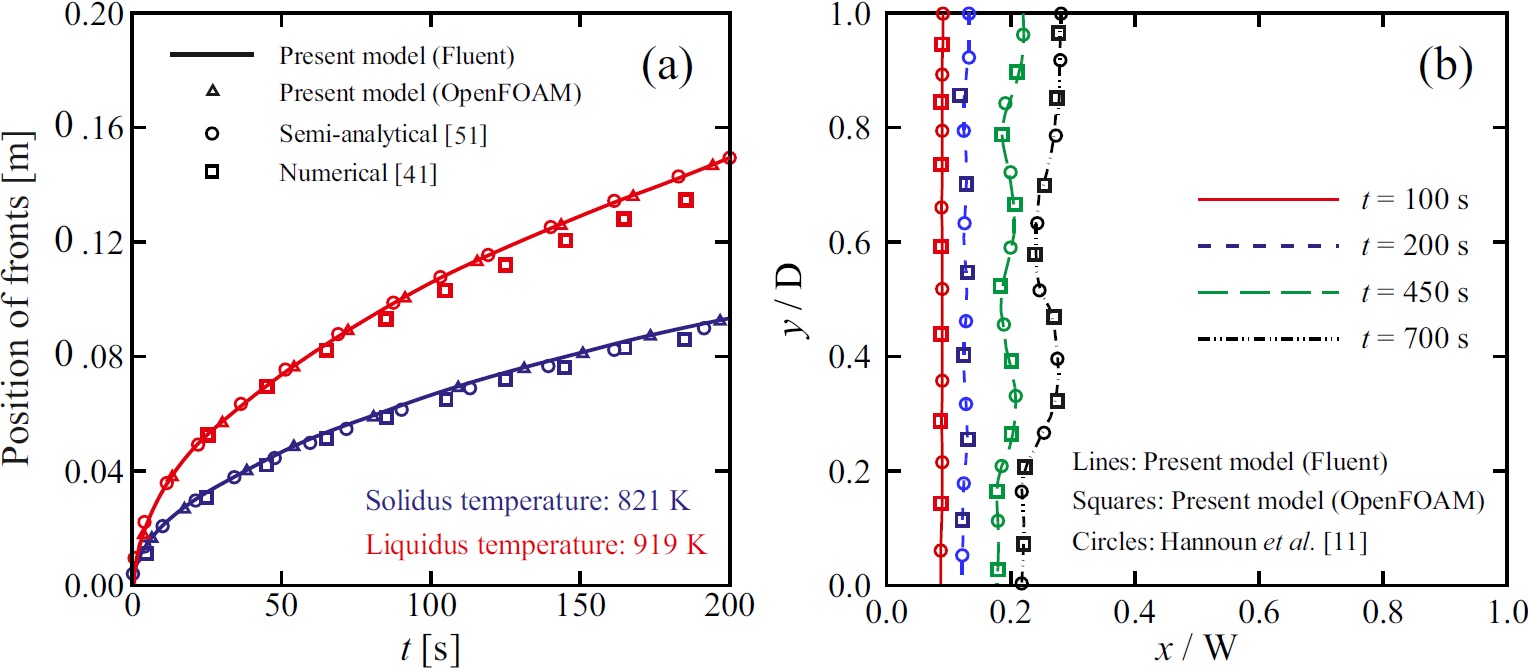}
	\caption{Code validation and solver verification. (\textbf{a}) One-dimensional benchmark problem; time evolution of the mushy-zone thickness (blue: solidus; red: liquidus) predicted by the present model implemented in Fluent and OpenFOAM compared with available semi-analytical~\cite{Voller_1989} and numerical~\cite{Voller_1991} results. (\textbf{b}) Two-dimensional benchmark problem; time evolution of solid--liquid interface during isothermal phase-change with convection predicted by the~present model implemented in Fluent and OpenFOAM compared with numerical results of Hannoun \textit{et al.}~\cite{Hannoun_2005}.}
	\label{fig: validation}
\end{figure}

\noindent
The reliability of the present numerical model was verified against available experimental, theoretical and numerical data for various phase-change benchmark problems. The~transient solidification of a~semi-infinite slab of Al--4.5\%Cu alloy was considered as a one-dimensional problem. The~slab is initially in the liquid phase with a uniform temperature of \SI{969}{\kelvin} above the liquidus temperature \mbox{$T_\mathrm{l}$ = \SI{919}{\kelvin}}. The~phase-change process starts at $t$ = \SI{0}{\second} by suddenly changing the temperature of one side of the slab to \SI{573}{\kelvin}, below~the solidus $T_\mathrm{s}$ = \SI{821}{\kelvin}. The~results of the present model on a uniform mesh with cell size $\Delta x =$ \SI{0.01}{\meter} are compared with semi-analytical~\cite{Voller_1989} and numerical results~\cite{Voller_1991} in \cref{fig: validation}a. Our present results are in better agreement with the semi-analytical solution from~\cite{Voller_1989} than the numerical results presented in~\cite{Voller_1991}, with~the maximum absolute {difference} between present results and the semi-analytical solution being 0.5\%.

Melting with convection of pure tin in a square enclosure was considered as a two-dimensional benchmark problem. Details of the problem can be found in~\cite{Hannoun_2005}. The~results obtained from the present model on a uniform mesh with quadrilateral grid cells, $\Delta x_\mathrm{i} / \mathrm{W} =$ \SI{4e-4}{} for isothermal phase-change including convection are compared with numerical results presented by \mbox{Hannoun \textit{et al.}~\cite{Hannoun_2005}} in \cref{fig: validation}b. The~maximum difference between the results predicted by the~present model and the reference case~\cite{Hannoun_2005} is within 1\%.

\begin{figure}[H]
	\centering
	\includegraphics[width=0.50\linewidth]{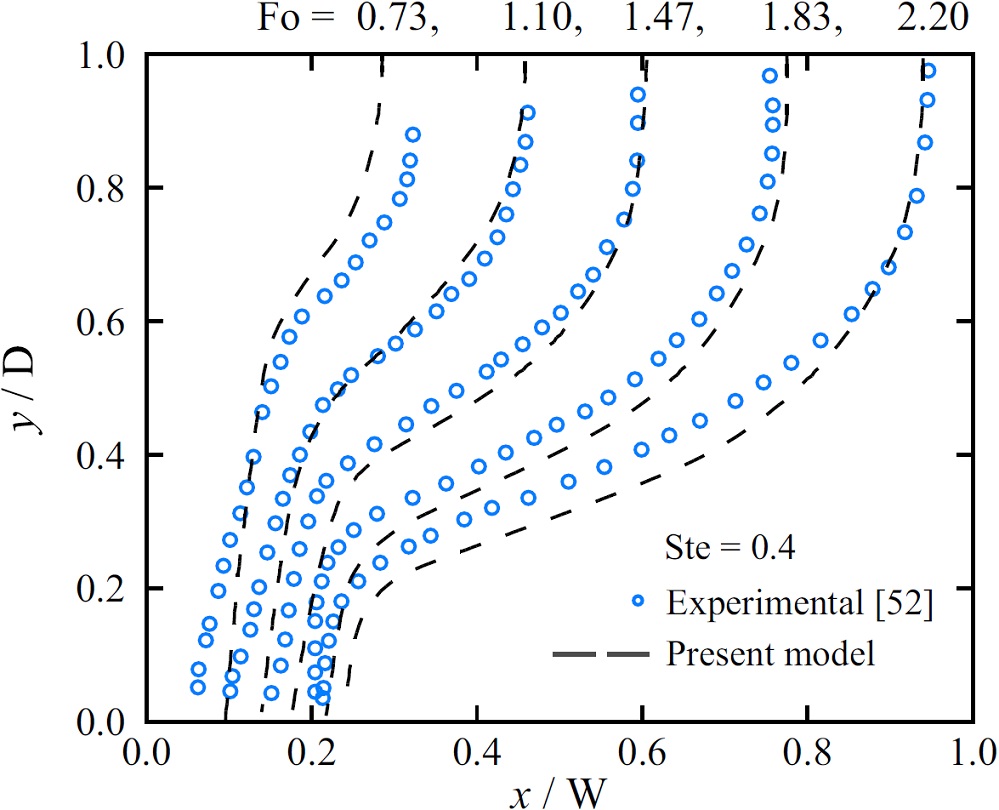}
	\caption{Comparisons of numerically determined melting-front positions in a rectangular enclosure subject to natural convection using the present model implemented in Fluent with experimental observations of isothermal melting of lead using thermal neutron radiography~\cite{Kumar_2012}.}
	\label{fig: validation-exp}
\end{figure}

Numerical predictions of solid--liquid interface morphologies in a rectangular enclosure subject to natural convection was compared with the experimental observations of Kumar \textit{et al.}~\cite{Kumar_2012} for isothermal melting of lead. \Cref{fig: validation-exp} indicates a reasonable agreement between the results obtained from the present numerical simulations and the reference data. The~deviations between the numerical and experimental data are attributed to both the simplifying assumptions made in the numerical model and the uncertainties associated with the experiments such as the uncertainty in determining the position of solid--liquid interface, which in this case is reported to be about 8.7\% \cite{Kumar_2012}. The~results obtained from the present model are also validated with experimental observations in \cref{sec: discussion} for a~laser spot melting process (see \cref{fig: weld_pool}).

\unskip

\subsection{Grid Size and Sensitivity to the Permeability Coefficient for Isothermal Phase~Change}
\label{sec: gridSensitivityIsothermal}

\noindent
Sensitivity of the solution to the grid size and the value of the permeability coefficient was studied for isothermal phase-change of gallium melting at $T_\mathrm{m}$ = \SI{302.78}{\kelvin}, in~a side-heated rectangular enclosure with $T_\mathrm{i} = T_\mathrm{c}$ = \SI{301.30}{\kelvin}, and~$T_\mathrm{h}$ = \SI{311.0}{\kelvin}, which was experimentally investigated by \mbox{Gau and Viskanta~\cite{Gau_1986}}. This benchmark case has often been considered for validation of phase-change simulations in the literature~\cite{Dutil_2011}. Detailed information regarding the benchmark case is available in~\cite{Gau_1986,Hannoun_2003}. 

\begin{figure}[H]
	\centering
	\includegraphics[width=1.00\linewidth]{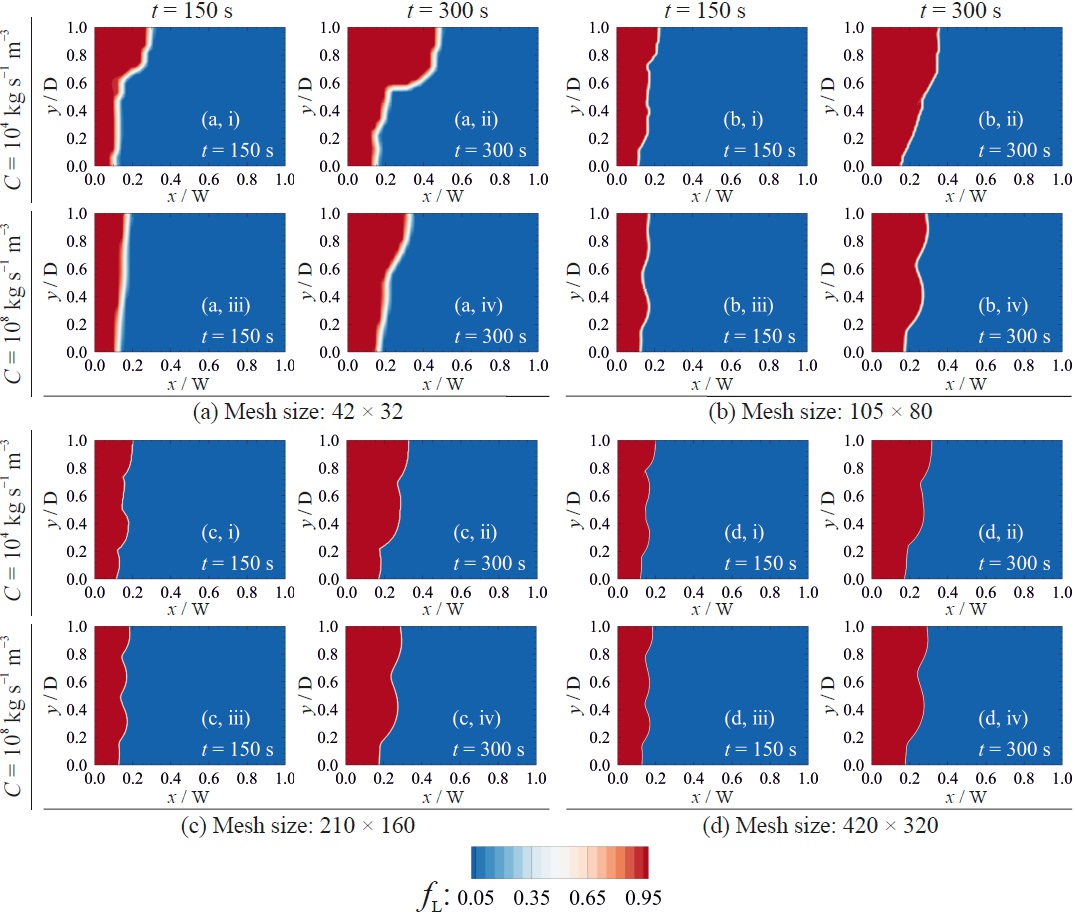}
	\caption{The influence of the computational cell size and the permeability coefficient on predicted melting front position for the gallium melting problem. Grid cell size: (\textbf{a}) $\Delta x / \mathrm{D} \approx \SI{3.33e-2}{}$; (\textbf{b})~$\Delta x / \mathrm{D} \approx \SI{1.33e-2}{}$; (\textbf{c})~$\Delta x / \mathrm{D} \approx \SI{6.67e-3}{}$; and (\textbf{d})~$\Delta x / \mathrm{D} \approx \SI{3.33e-3}{}$.}
	\label{fig: c_gridstudy}
\end{figure}

\begin{figure}[H]
	\centering
	\includegraphics[height=0.300\textheight]{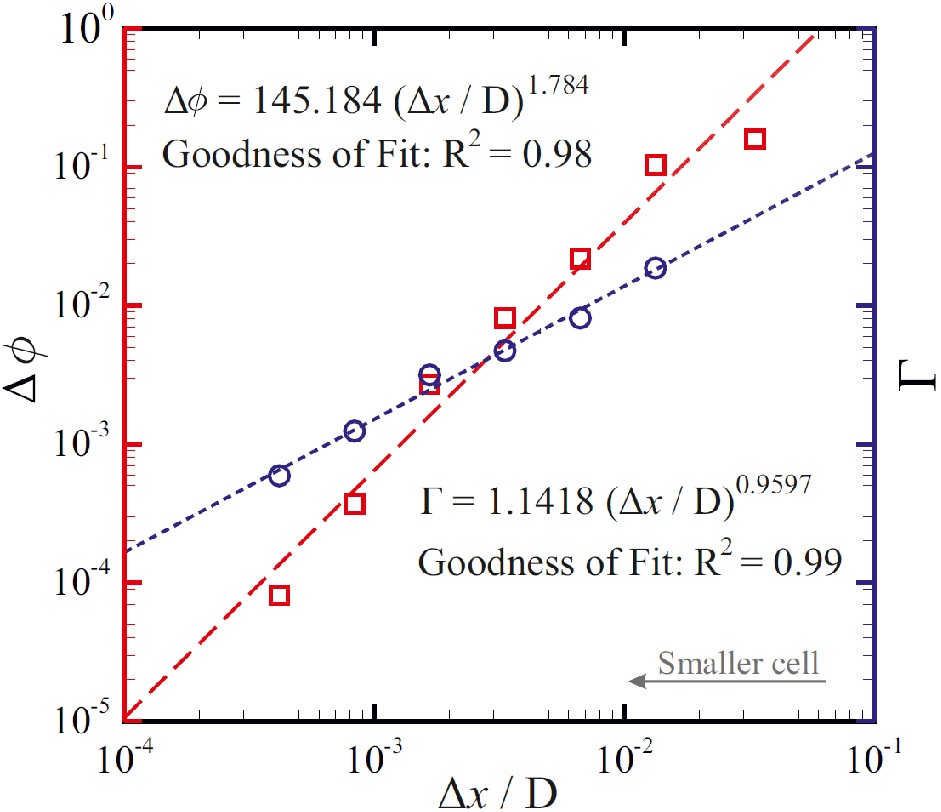}
	\caption{The relative difference between the solid--liquid interface morphologies ($\Gamma$ (defined in \cref{eq: gamma}), circles in blue), and~the relative difference between the liquid fractions ($\Delta\phi$~(defined below \cref{eq: phi}), squares in red) at $t = \SI{150}{\second}$ when using permeability coefficients $C_1 = \SI{1e8}{\kilogram\per\second\per\meter\cubed}$ and $C_2 = \SI{1e4}{\kilogram\per\second\per\meter\cubed}$ as a function of grid size ($\Delta x$). Larger values of $\Delta\phi$ and $\Gamma$ indicate more sensitivity to the value of the permeability coefficient $C$. Symbols: results obtained from numerical simulations; dashed lines: curve-fit.}
	\label{fig: quantification_gridstudy}
\end{figure}

\begin{figure}[H]
	\centering
	\includegraphics[height=0.30\textheight]{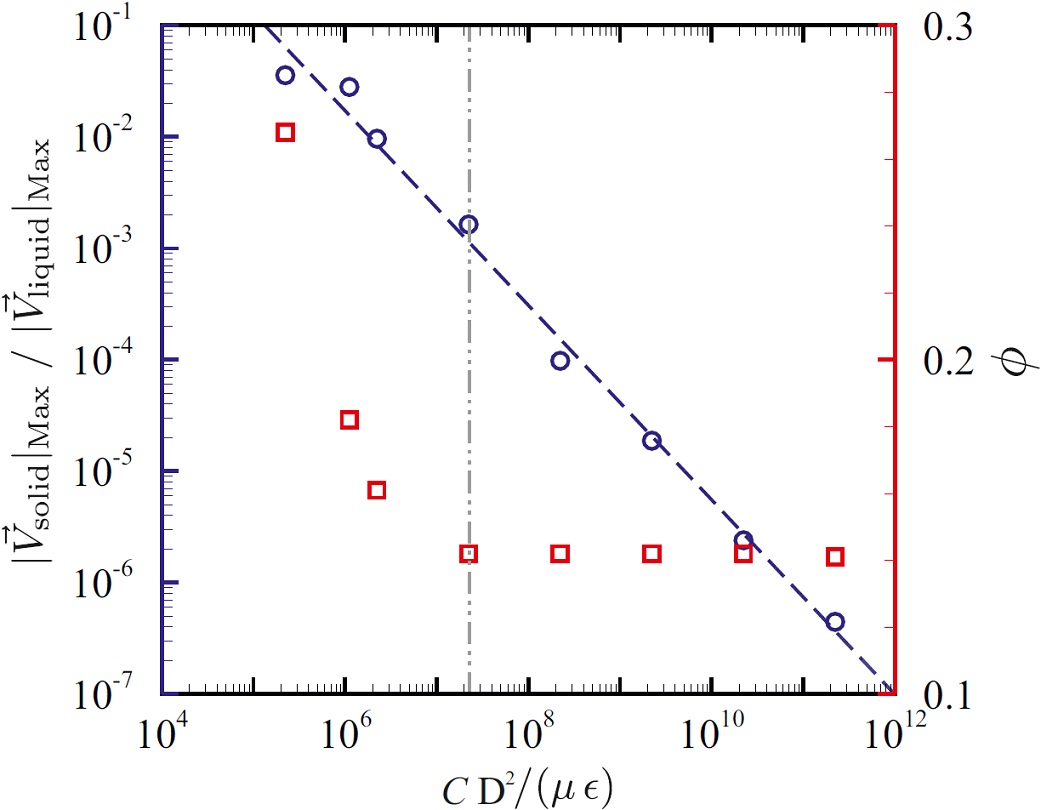}
	\caption{The ratio of the maximum velocity magnitude in solid regions to that in liquid regions ($|\vec{V}_{\mathrm{solid}}|_\mathrm{Max} \ / \  |\vec{V}_{\mathrm{liquid}}|_\mathrm{Max}$, blue circles) and the predicted liquid fractions $\phi$ (red squares) for various values of the sink term $\left(C\,\mathrm{D}^2 /\left(\mu\epsilon\right)\right)$.}
	\label{fig: SolidVelocity_SinkTerm}
\end{figure}

\Cref{fig: c_gridstudy} shows the influence of the permeability coefficient on the predicted liquid fraction, using uniform fixed meshes with 42 $\times$ 32, 105 $\times$ 80, 210 $\times$ 160 and 420 $\times$ 320 computational grid cells and $C = 10^4$ and $10^8\,\SI{}{\kilogram\per\second\per\meter\cubed}$. When a coarse mesh is utilised, resulting in a non-physical mushy zone of significant thickness, the~solution appears to be very sensitive to the value of the~permeability coefficient. This sensitivity is attributed to the enhanced heat transfer from the~hot~fluid to the solid--liquid interface due to higher fluid velocities in the mushy zone predicted with a smaller permeability coefficient. Reducing the grid cell size, and~thus reducing the thickness of the non-physical mushy zone, decreases the total volume of the grid cell in which the permeability coefficient affects the numerical predictions. Consequently, the~sensitivity of the solution to the~value of the~permeability coefficient decreases with grid refinement. A~finite amount of time is required for the mass in a computational cell to absorb heat and melt. A~change in the value of $C$ therefore affects the convective heat transfer to the cells located at the solid--liquid interface that can lead to a change in the predicted interface morphology and the rate of melting during the transient phase. This effect reduces by refining the grid size adjacent to the solid--liquid interface~\cite{Mencinger_2004}. In~\cref{fig: quantification_gridstudy}, the~sensitivity to the value of $C$ is quantified as a function of grid cell size by looking at the~parameters $\Delta\phi$ and $\Gamma$ (defined below \cref{eq: phi} and in \cref{eq: gamma}, representing the relative difference in the predicted liquid fractions, and~the relative difference between the~solid--liquid interface morphologies) for $C_1 = \SI{1e8}{\kilogram\per\second\per\meter\cubed}$ and $C_2 = \SI{1e4}{\kilogram\per\second\per\meter\cubed}$ at $t = \SI{150}{\second}$. Reducing the cell size decreases the influence of the permeability coefficient on the results, with~$\Gamma$ scaling approximately linearly with cell size, and~$\Delta\phi$ scaling approximately quadratically with cell size. Thus, although~in principle the chosen value of $C$ should be irrelevant for isothermal phase change, sufficient grid refinement is needed to obtain solutions which are indeed insensitive to the~value of $C$. In~addition, sufficient grid resolution in the liquid zone was found to be required to predict the multicellular convection~\cite{Lee_1983,Dantzig_1989,Le_Quere_1999,Cerimele_2002} early in the phase-change process, which leads to the formation of a wavy solid--liquid interface. According to \cref{eq: sink_term_voller}, the~magnitude of the sink-term in solid regions ($f_\mathrm{L} = 0.0$) should be large compared to the viscous term ($\nabla^2\vec{V}^*$) to suppress fluid velocities (\textit{i.e.}, $C\,\mathrm{D}^2 /\left(\mu\epsilon\right) \gg 1$). The~influence of the sink-term magnitude on predicted liquid fraction $\phi$ and the ratio of the maximum velocity magnitude in solid regions to that in liquid regions (\textit{i.e.},~$|\vec{V}_{\mathrm{solid}}|_\mathrm{Max} \ / \  |\vec{V}_{\mathrm{liquid}}|_\mathrm{Max}$) is shown in \cref{fig: SolidVelocity_SinkTerm}. Here, $\mathrm{D}$ is chosen to be the~size of the heated wall. For~magnitudes of $C\,\mathrm{D}^2 /\left(\mu\epsilon\right)$ roughly larger than $\SI{2.5e7}{}$, the~predicted liquid fraction is independent of the sink-term. For~smaller values of the sink-term, even though the velocity magnitude in the solid region is orders of magnitude smaller than that in the liquid region, the~energy transfer to the solid material is affected, which leads to a change in the predicted liquid~fraction.

The fine-grid results presented in \cref{fig: c_gridstudy} represent a mathematically converged solution of the~governing equations for the given set of boundary conditions. However, the~grid independent results of this particular case do not closely match with the experimental results reported in~\cite{Gau_1986}. Obtaining a better agreement between numerical predictions and experimental observations by using a coarser grid (\textit{i.e.}, grid-dependent results), or~by tuning the mathematical model has been reported in the~literature~\cite{Kumar_2017} that is coincidental, but~such \textit{ad~hoc} tuning lacks generality~\cite{Hannoun_2003,Hannoun_2005,Schroeder_2017}.

\unskip

\subsection{Non-Isothermal~Phase-Change}
\label{sec: nonisothermalPhaseChange}
\vspace{-6pt}

\subsubsection{Grid~Sensitivity}
\label{sec: gridSensitivityNonisothermal}

\begin{figure}[H]
	\centering
	\includegraphics[height=0.300\textheight]{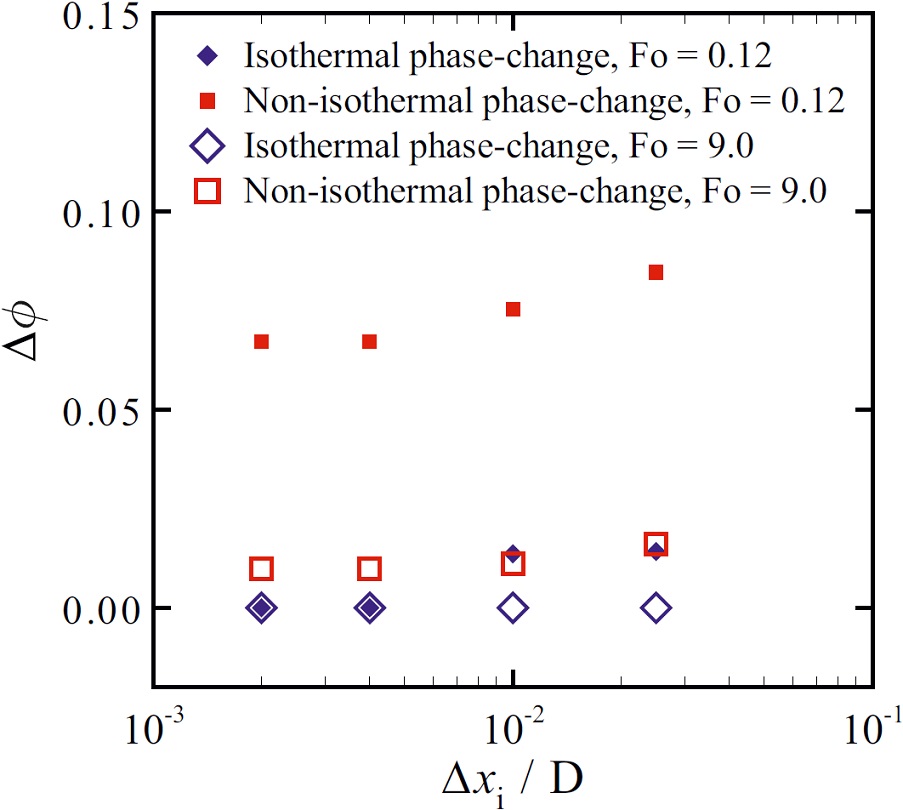}
	\caption{The relative difference between liquid fractions predicted with different permeability coefficients (\mbox{\textit{i.e.},~$C_1 = \SI{1e8}{\kilogram\per\second\per\meter\cubed}$} and $C_2 = \SI{1e4}{\kilogram\per\second\per\meter\cubed}$) on different grid densities. Diamonds (in blue): isothermal phase-change; squares (in red): non-isothermal phase-change; unfilled symbols: Fo = 9.0 (steady-state condition); filled symbols: Fo = 0.12.}
	\label{fig: liquidFraction_gridStudy}
\end{figure}

\noindent
Having seen the importance of sufficient grid resolution for isothermal phase-change, we~considered the non-isothermal phase-change problem defined in \cref{sec: problemDescription}, examining the impact of grid refinement on the sensitivity to the permeability coefficient $C$. For~$\Delta T_\mathrm{m} = \SI{50}{\kelvin}$, \cref{fig: liquidFraction_gridStudy} shows the sensitivity parameter $\Delta\phi$, for~short time (Fo = 0.12) and steady-state (Fo = 9.0) obtained with $C_1 = \SI{1e8}{\kilogram\per\second\per\meter\cubed}$ and $C_2 = \SI{1e4}{\kilogram\per\second\per\meter\cubed}$. For~comparison, the~same problem was also solved assuming isothermal phase change ($\Delta T_\mathrm{m} = \SI{0}{\kelvin}$). For~isothermal phase change, the~results are insensitive to the chosen value of $C$ in the steady-state, and~exhibit a small sensitivity to the chosen value of $C$ during the transient phase before reaching the steady-state that becomes negligible with reducing grid size. For~non-isothermal phase change, the~sensitivity to the chosen value of $C$ is much stronger and does not vanish with decreasing grid size, even in the steady-state, as~is to be expected from the fact that a physically realistic mushy zone is now present, in~which flow and convective heat transfer are sensitive to the value of $C$. However, the~sensitivity to $C$ becomes grid independent for base grid sizes below $\Delta x_\mathrm{i} / \mathrm{D} = \SI{4e-3}{}$. Consequently, this base grid size is used in the next sections. In~addition, a~four-level dynamic solution-adaptive mesh refinement (\textit{i.e.},~$\Delta x_\mathrm{N} = \Delta x_\mathrm{i} / 2^{\mathrm{N}}, \mathrm{N} = 1, \cdots,4$) is applied to further enhance the accuracy with which we capture the solid--liquid~interface.

\unskip

\subsubsection{Influence of the Permeability Coefficient on Predicted~Results}
\label{sec: permeabilityInfluenceNonisothermal}

\begin{figure}[H]
	\centering
	\includegraphics[height=0.23\textheight]{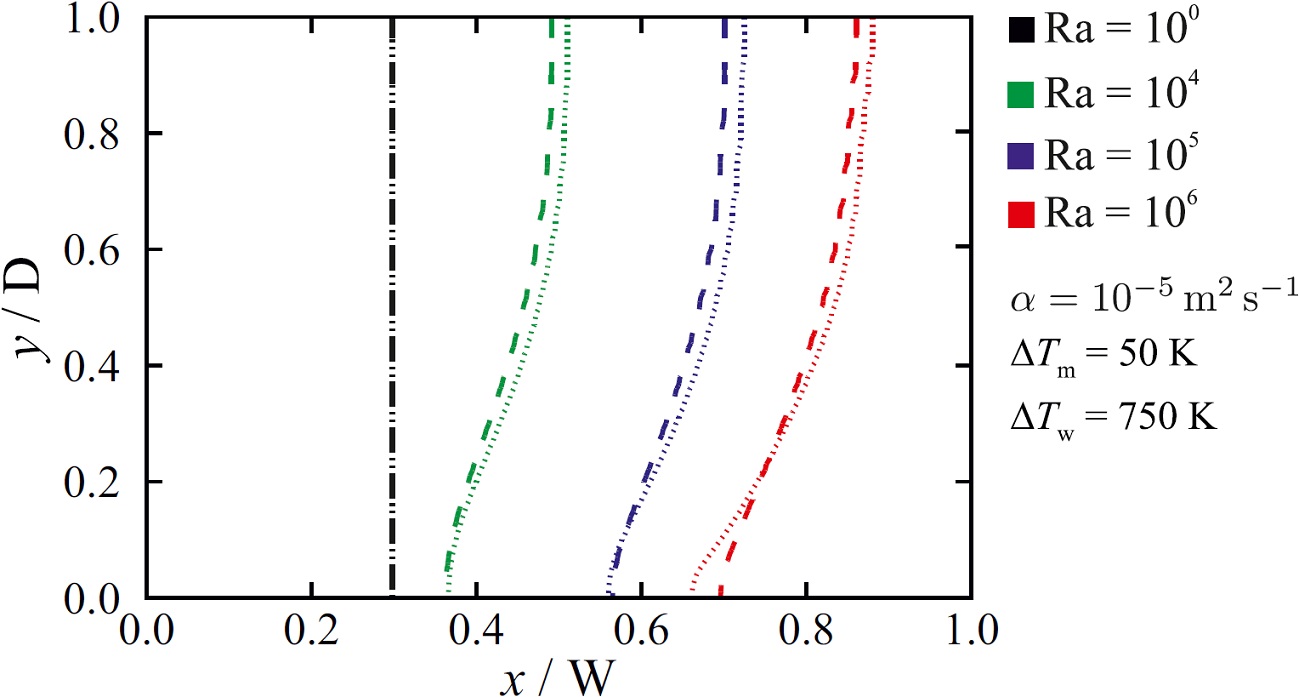}
	\caption{The influence of the permeability coefficient on predicted liquidus-line ($f_\mathrm{L}$ = 1) position for different Rayleigh numbers ($\Delta T_\mathrm{m} = \SI{50}{\kelvin}$; $\Delta T_\mathrm{w} = \SI{750}{\kelvin}$; dashed lines: $C = \SI{1e8}{\kilogram\per\second\per\meter\cubed}$; dotted lines: $C = \SI{1e4}{\kilogram\per\second\per\meter\cubed}$). The~results are given for Fo = 9, in~which steady-state solutions are~achieved.}
	\label{fig: liquidus_line_Ra}
\end{figure}

\noindent
For the non-isothermal phase change problem defined in \cref{sec: problemDescription}, the~influence of the permeability coefficient on the predicted steady-state liquidus-line position ($f_\mathrm{L} = 1$) is shown in \cref{fig: liquidus_line_Ra} for a~fixed wall temperature difference ($\Delta T_\mathrm{w}$), a~fixed melting temperature range ($\Delta T_\mathrm{m}$) and varying Rayleigh number (\textit{i.e.}, varying fluid velocities). In~the virtual absence of flow, for~Ra = 1, the~results are insensitive to the value of the permeability coefficient. In~this case, heat conduction dominates the~total energy transfer around and in the mushy zone. By~increasing the value of Ra, \textit{i.e.}~increasing fluid flow velocities and increasing convective heat transfer, the~results become more sensitive to the~chosen value of $C$. This can be understood from the fact that the value of $C$, through the~momentum sink term, only influences the convective terms and therefore the results become more sensitive to $C$ when convection plays an important role in total heat transfer in the mushy~zone.

\begin{figure}[H]
	\centering
	\includegraphics[height=0.23\textheight]{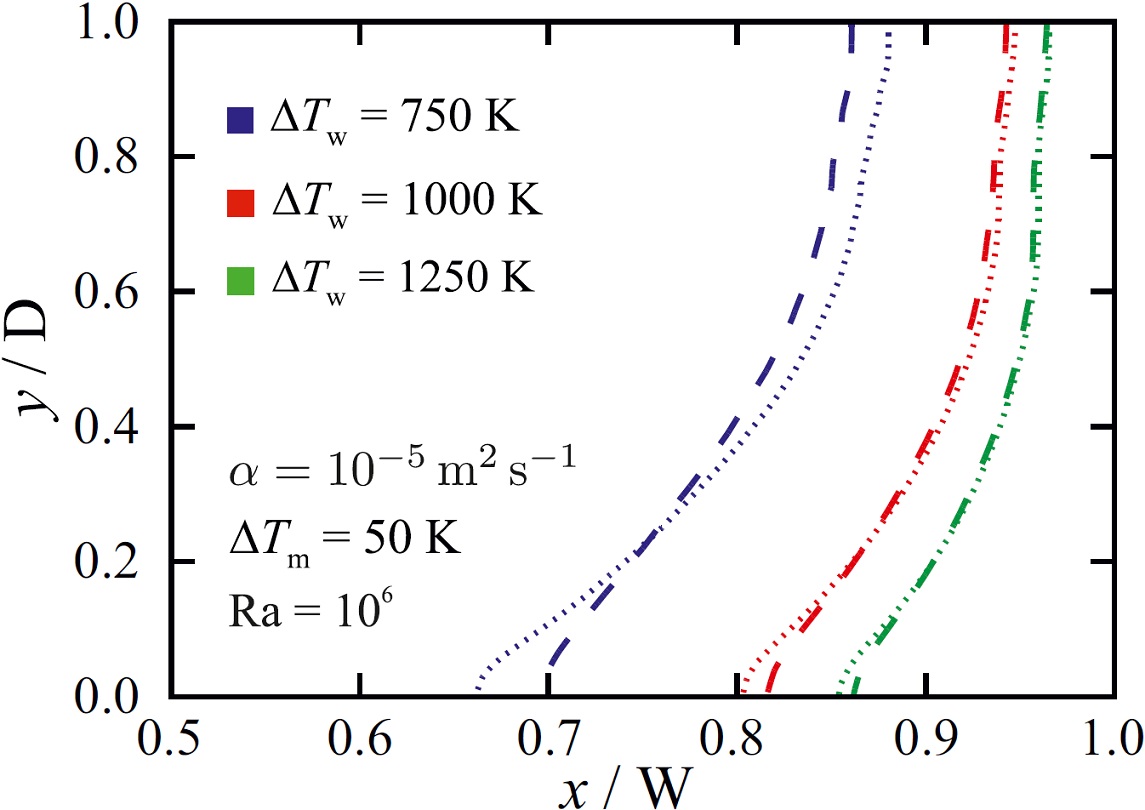}
	\caption{The influence of the permeability coefficient on predicted liquidus-line ($f_\mathrm{L}$ = 1) position for different wall-temperature differences $\Delta T_\mathrm{w}$, at~$\Delta T_\mathrm{m} = \SI{50}{\kelvin}$, and~Ra = $10^6$. Dashed lines: $C = \SI{1e8}{\kilogram\per\second\per\meter\cubed}$; dotted lines: $C = \SI{1e4}{\kilogram\per\second\per\meter\cubed}$. The~results are given for Fo = 9, in~which steady-state solutions are~achieved.}
	\label{fig: liquidus_line_Tw}
\end{figure}

\begin{figure}[H]
	\centering
	\includegraphics[height=0.23\textheight]{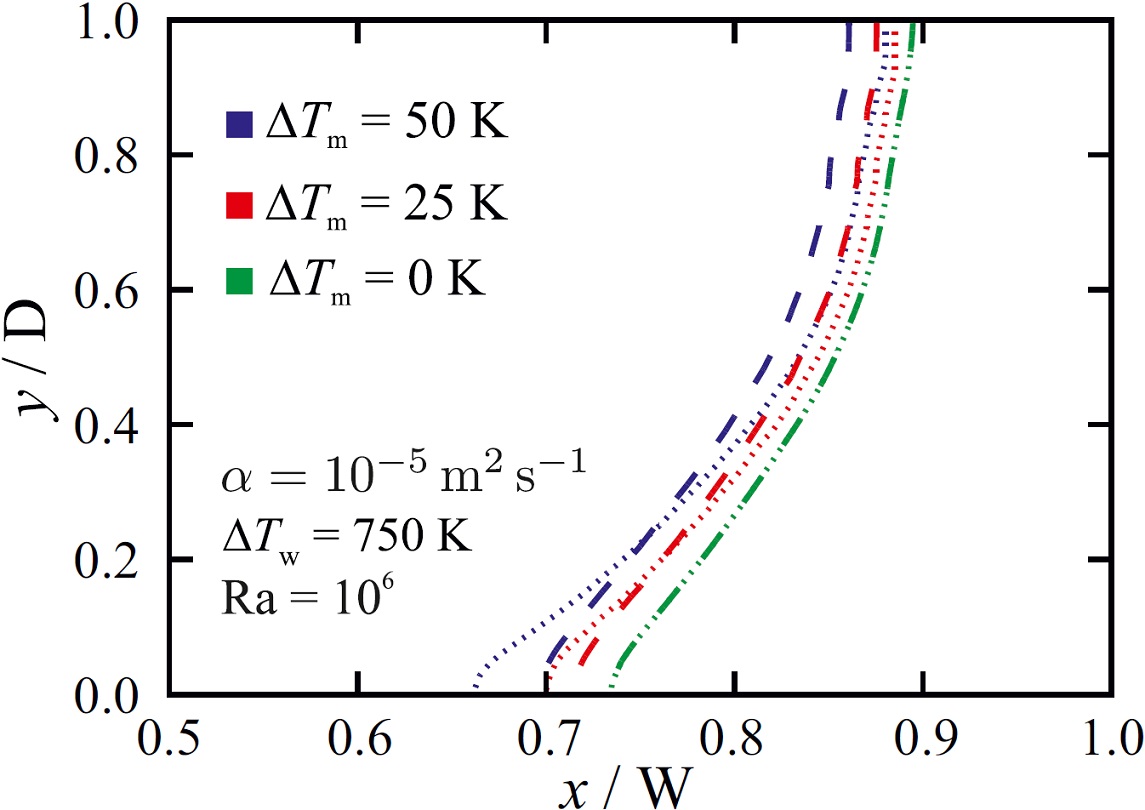}
	\caption{The influence of the permeability coefficient on predicted liquidus-line ($f_\mathrm{L}$ = 1) position for different melting-temperature ranges $\Delta T_\mathrm{m}$, at~$\Delta T_\mathrm{w} = \SI{750}{\kelvin}$, and~Ra = $10^6$. Dashed lines: $C = \SI{1e8}{\kilogram\per\second\per\meter\cubed}$; dotted lines: $C = \SI{1e4}{\kilogram\per\second\per\meter\cubed}$. The~results are given for Fo = 9, in~which steady-state solutions are~achieved.}
	\label{fig: liquidus_line_Tm}
\end{figure}

\begin{figure}[H]
	\centering
	\includegraphics[height=0.23\textheight]{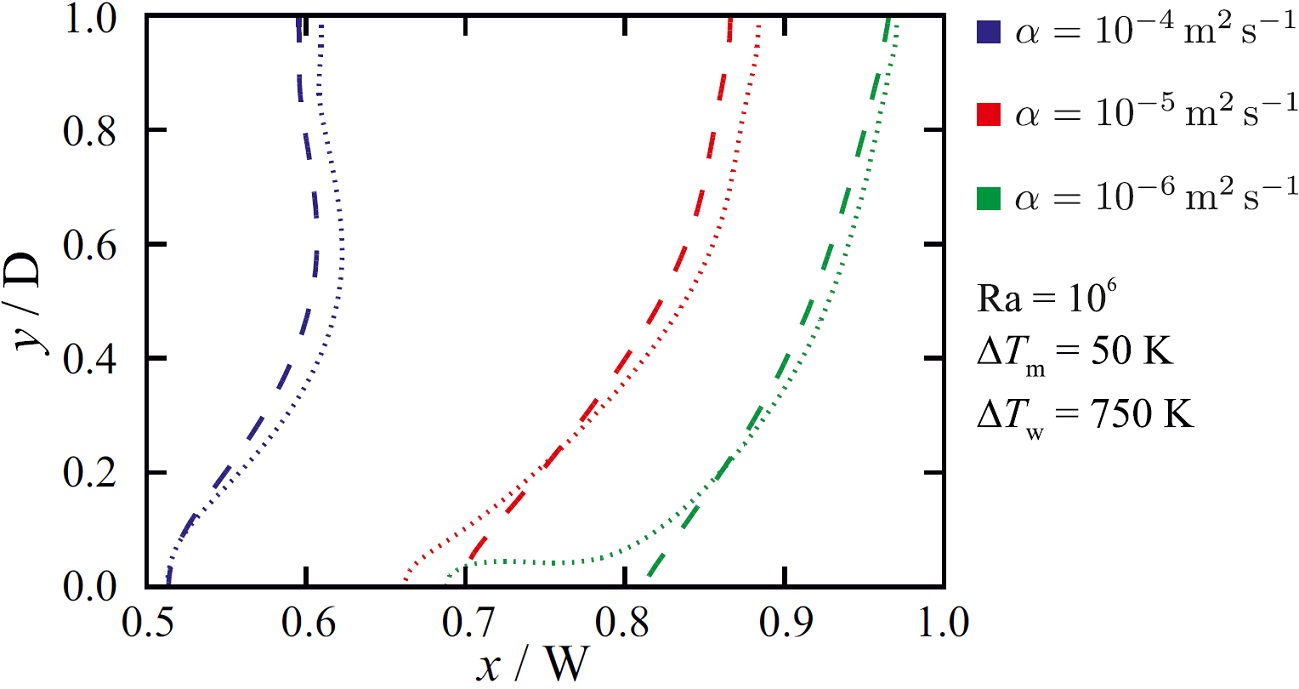}
	\caption{The influence of the permeability coefficient on predicted liquidus-line ($f_\mathrm{L}$ = 1) position for different values of thermal diffusivity $\alpha$, at~$\Delta T_\mathrm{m} = \SI{50}{\kelvin}$, $\Delta T_\mathrm{w} = \SI{750}{\kelvin}$ and Ra = $10^6$. Dashed lines: $C = \SI{1e8}{\kilogram\per\second\per\meter\cubed}$; dotted lines: $C = \SI{1e4}{\kilogram\per\second\per\meter\cubed}$. The~results are given for Fo = 9, in~which steady-state solutions are~achieved.}
	\label{fig: liquidus_line_Alpha}
\end{figure}

\Cref{fig: liquidus_line_Tw} {shows} the effect of the permeability coefficient for a fixed melting temperature range ($\Delta T_\mathrm{m}$) with different wall-temperature differences ($\Delta T_\mathrm{w}$). Here, a~higher value of $\Delta T_\mathrm{w}$ (\textit{i.e.}, smaller $\theta$) leads to a smaller mushy zone thickness, and~consequently a reduced sensitivity to the permeability coefficient. Similarly, for~a fixed ($\Delta T_\mathrm{w}$), reducing ($\Delta T_\mathrm{m}$) decreases the mushy zone thickness and as a result lower the sensitivity to the permeability coefficient $C$, as~shown in \cref{fig: liquidus_line_Tm}. In~summary, the~results show that non-isothermal phase-change simulations are less affected by the chosen value of the permeability coefficient $C$ when the mushy zone has a smaller thickness (\textit{i.e.}, smaller $\theta$), as~this leads to the conductive heat transfer through the mushy zone being large compared to the convective heat transfer. A~change in the thermal diffusivity of the material can therefore affect the sensitivity of the numerical predictions to the value of the permeability coefficient. \Cref{fig: liquidus_line_Alpha} indicates the sensitivity of the results to the chosen value of the permeability coefficient as a function of thermal diffusivity of the material for a fixed melting temperature range ($\Delta T_\mathrm{m}$) and wall-temperature difference ($\Delta T_\mathrm{w}$). It is seen that sensitivity to the permeability coefficient decreases with increasing thermal diffusivity of the material, which can be attributed to the enhancement of heat conduction contribution to the total heat~transfer.

\unskip

\unskip

\begin{figure}[H]
	\centering
	\includegraphics[width=1.00\linewidth]{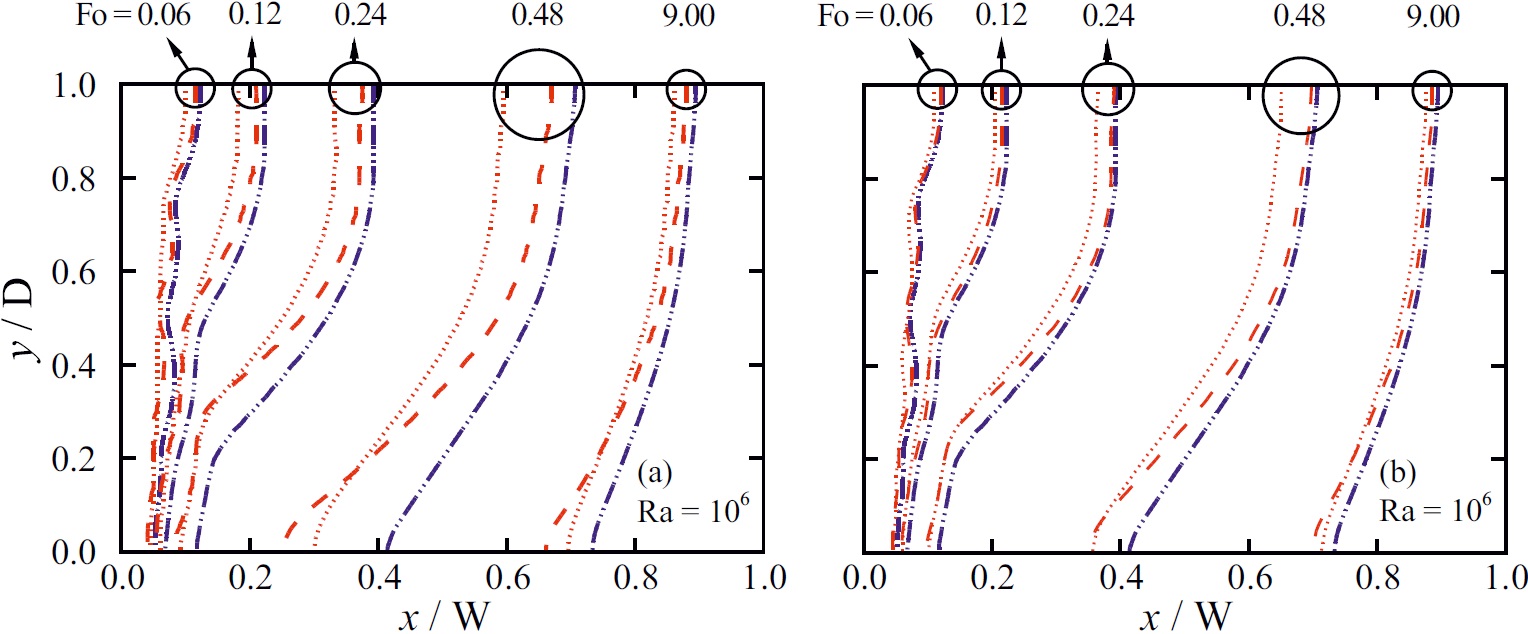}
	\caption{Time evolution of melting front positions ($f_\mathrm{L}$ = 1) for isothermal phase-change (blue lines, $\theta$ = 0) and non-isothermal phase-change (red lines: (\textbf{a}) $\Delta T_\mathrm{m} = \SI{50}{\kelvin}$, $\Delta T_\mathrm{w} = \SI{750}{\kelvin}$, $\theta$ = 1/15; and (\textbf{b}) $\Delta T_\mathrm{m} = \SI{25}{\kelvin}$, $\Delta T_\mathrm{w} = \SI{750}{\kelvin}$, $\theta$ = 1/30) problems. Dashed lines: $C = \SI{1e4}{\kilogram\per\second\per\meter\cubed}$; dotted lines: $C = \SI{1e8}{\kilogram\per\second\per\meter\cubed}$  (Ra = $10^6$).}
	\label{fig: time_evolution}
\end{figure}

The influence of the permeability coefficient on the time evolution of the melt pool shape is presented in \cref{fig: time_evolution}. The~results are indeed independent of the permeability coefficient for an~isothermal phase-change. A~similar conclusion has also been drawn after monitoring the time variations of the liquid fraction obtained from numerical simulations~\cite{Vogel_2019,Kumar_2012}. However, for~non-isothermal phase change with a thick mushy zone ($\theta$ = 1/15) and strong flow (Ra = $10^6$), the~results are very sensitive to the chosen value of the permeability coefficient, resulting is different pool shapes and rates of phase-change. Less sensitivity to the value of $C$ is found for a thin mushy zone ($\theta = 1/30$). The~numerical predictions are also less sensitive to the permeability coefficient at early time instances, when the fluid flow is characterised by low~velocities. 

\unskip

\section{Discussion}
\label{sec: discussion}

\noindent
The results reveal that, depending on the temperature gradient, velocity field and thermophysical properties of the phase-change material, which in turn determine the mushy zone thickness, numerical predictions of phase-change problems can show sensitivity to the value of the permeability coefficient. A~general guideline is presented here that allows prediction and evaluation of the influence of the~permeability coefficient in phase-change simulations. This concept involves both the heat transfer mechanism and the mushy zone~thickness.

For phase-change problems without fluid flow, the~mushy zone thickness ($m_\mathrm{t}$) can be estimated as
\begin{equation}
m_\mathrm{t} = \frac{\Delta T_\mathrm{m}}{\left|\nabla T\right|} = 	\frac{\Delta T_\mathrm{m} \cdot \mathscr{L}}{T_\mathrm{h} - T_\mathrm{c}},
\label{eq: mushy_thinkness}
\end{equation}

\noindent where $\mathscr{L}$ is a characteristic length scale. The~fluid flow can alter the mushy zone thickness when convective heat transfer in the mushy zone is of significance. This can be expressed through a~P{\'e}clet~number, which expresses the ratio between the rate of heat advection and heat diffusion. To~identify the regions where there is a significant heat transfer enhancement or reduction due to convection, and~consequently where the results are sensitive to the permeability coefficient, $\mathrm{Pe^*}$ is defined as~follows:
\begin{equation}
\mathrm{Pe^*} = \frac{|\vec{V}| \cdot m_\mathrm{t}}{\alpha} = \frac{|\vec{V}| \cdot \Delta T_\mathrm{m} \cdot \mathscr{L}}{\alpha \left(T_\mathrm{h} - T_\mathrm{c}\right)}.
\label{eq: dimensionless_number}
\end{equation}

Non-zero values of $\mathrm{Pe^*}$ adjacent to the solid--liquid interface indicate increased sensitivity to the permeability coefficient. For~isothermal phase-change, $\mathrm{Pe^*}$ is zero and predictions are independent of the~permeability~coefficient.

\begin{figure}[H]
	\centering
	\includegraphics[width=1.00\linewidth]{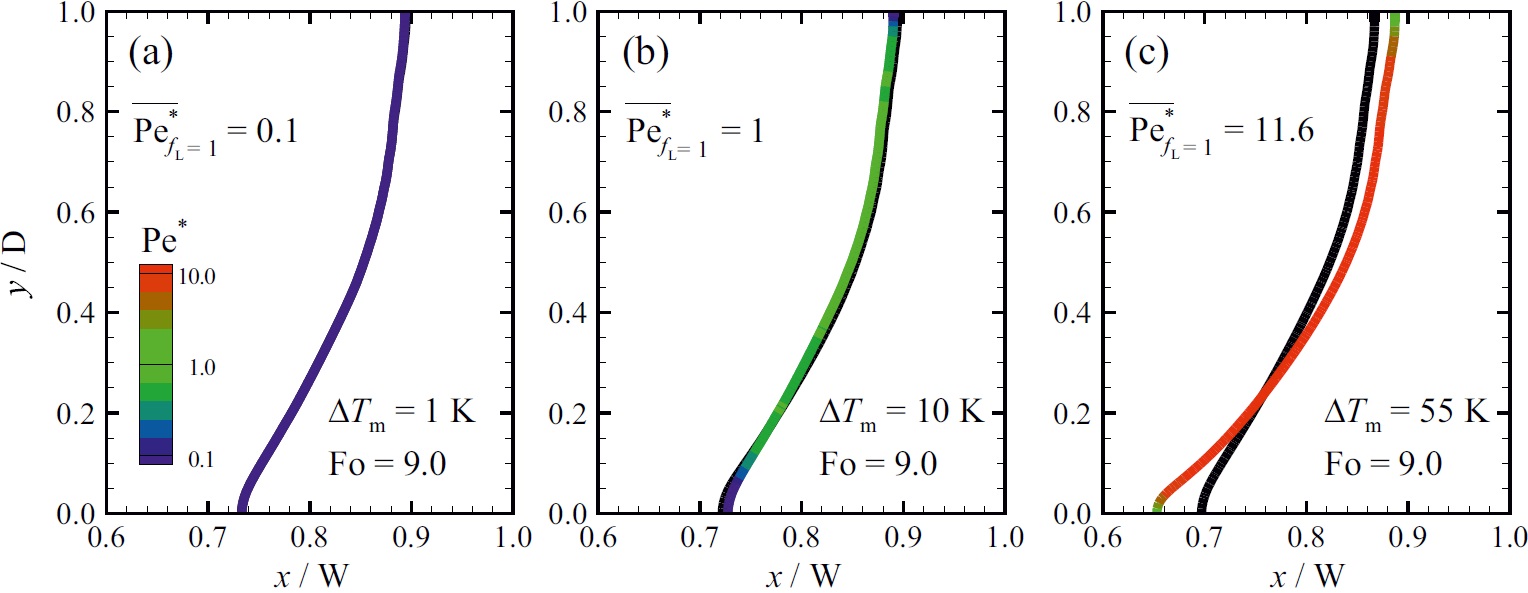}
	\caption{Melting front positions ($f_\mathrm{L}$ = 1) for the problem described in \cref{sec: problemDescription}  ($\Delta T_\mathrm{w} = \SI{750}{\kelvin}$ and Ra = $10^6$) at the steady-state condition (Fo = 9) and values of $\mathrm{Pe^*}$ along the melting fronts for: (\textbf{a})~$\Delta T_\mathrm{m} = \SI{1}{\kelvin}$; (\textbf{b}) $\Delta T_\mathrm{m} = \SI{10}{\kelvin}$; and (\textbf{c}) $\Delta T_\mathrm{m} = \SI{55}{\kelvin}$ predicted using different permeability coefficients of $C = 10^4\,\SI{}{\kilogram\per\second\per\meter\cubed}$ (coloured lines), and~$10^{8} \,\SI{}{\kilogram\per\second\per\meter\cubed}$ (black lines). Values of $\mathrm{Pe^*}$ along the melting front are $\mathcal{O}(0.1)$, $\mathcal{O}(1)$ and $\mathcal{O}(10)$ for $\Delta T_\mathrm{m} = \SI{1}{\kelvin}$, $\SI{10}{\kelvin}$ and $\SI{55}{\kelvin}$, respectively. The~results are more sensitive to the permeability coefficient for large $\mathrm{Pe^*}$. For~the case with $\Delta T_\mathrm{m} = \SI{1}{\kelvin}$ (a), predictions with different permeability coefficient are identical and therefore the coloured line covers the black~line.}
	\label{fig: pe_buoyancy}
\end{figure}

Sensitivity of the numerical predictions to the value of permeability coefficient $C$ has been appraised for the problem defined in \cref{sec: problemDescription}, using three different melting temperature ranges $\Delta T_\mathrm{m}$ and a fixed $\Delta T_\mathrm{w} = \SI{750}{\kelvin}$, and~the results are shown in \cref{fig: pe_buoyancy}. Higher values of $\mathrm{Pe^*}$~($\approx \mathcal{O}(10)$) along the melting front for the case with $\Delta T_\mathrm{m} = \SI{55}{\kelvin}$ (\cref{fig: pe_buoyancy}c) compared to the case with $\Delta T_\mathrm{m} = \SI{10}{\kelvin}$ (\cref{fig: pe_buoyancy}b) indicate that the numerical predictions are more sensitive to the permeability coefficient, which is consistent with the results presented in \cref{fig: liquidus_line_Tm,fig: time_evolution}. When the values of $\mathrm{Pe^*}$ along the interface are small (\textit{i.e.},    $\mathrm{Pe^*} \ll 1$), the~results appear to be insensitive to the permeability coefficient, while, for~$\mathrm{Pe^*} \gg 1$, the~results are sensitive to $C$.

The general applicability of the proposed $\mathrm{Pe^*}$ criterion is also examined for a simulation of a~laser spot melting process, which was carried out experimentally by Pitschender \textit{et al.}~\cite{Pitscheneder_1996}. A~steel plate containing $\SI{20}{ppm}$ of sulphur was heated using a stationary laser beam with a power of $\SI{5200}{\watt}$ and a~top-hat radius of $\SI{1.4}{\milli\meter}$. The~surface absorptivity is set to 0.13~\cite{Pitscheneder_1996}. Material properties of the steel plate are assumed to be constant and temperature independent, except~for the surface tension of the molten material, and~can be found in~\cite{Pitscheneder_1996,Yan_2018,Saldi_2013}. Variations of surface tension with temperature are modelled using the expression proposed by Sahoo \textit{et al.}~\cite{Sahoo_1988}. Melting of the~material and associated heat and fluid flow are numerically predicted using permeability coefficients of $C = 10^6$, $10^8$ and $10^{10}\, \SI{}{\kilogram\per\second\per\meter\cubed}$. To~study the influence of the mushy-zone thickness on the sensitivity to $C$ through $\mathrm{Pe^*}$, the~melting-temperature range ($\Delta T_\mathrm{m}$) of the material is changed artificially to $\SI{40}{\kelvin}$ and $\SI{200}{\kelvin}$.

\begin{figure}[H]
	\centering
	\includegraphics[width=0.85\linewidth]{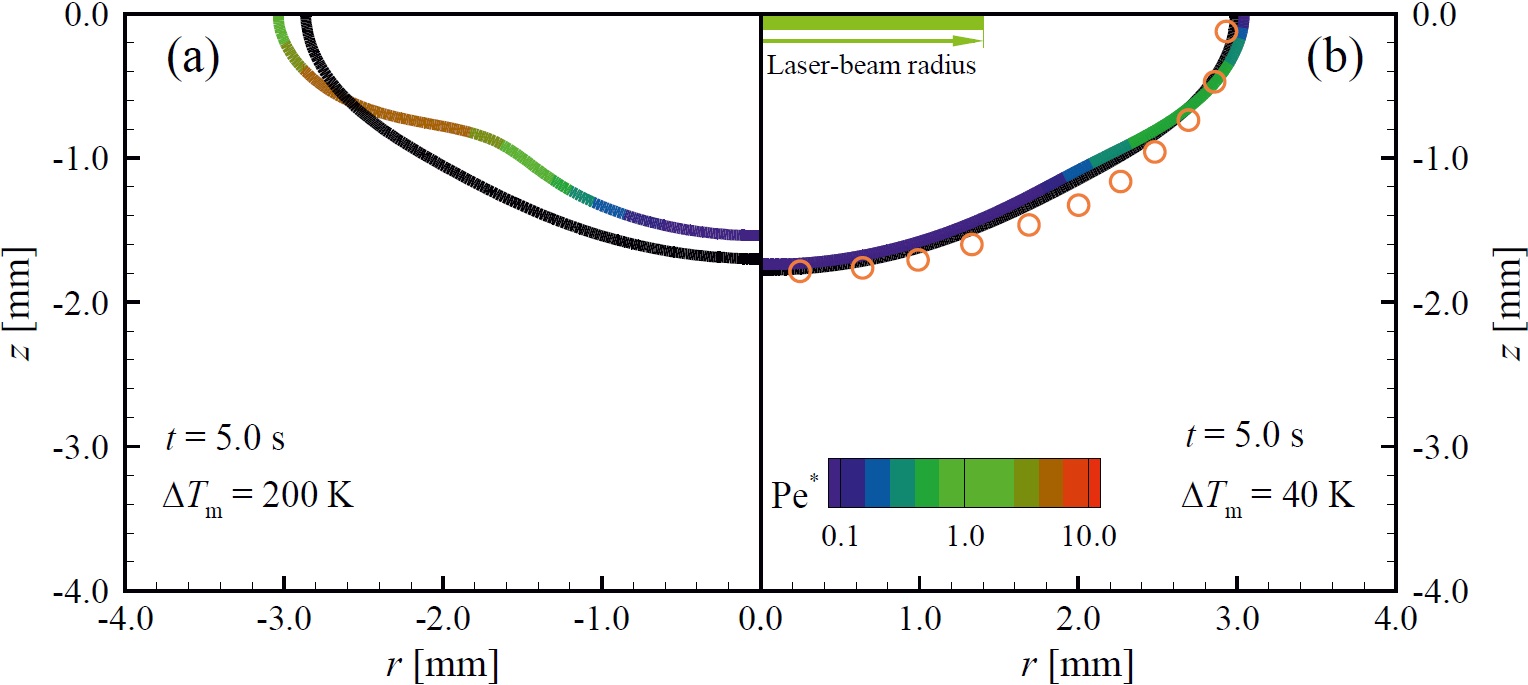}
	\caption{Laser spot melting of Fe-S binary alloys with $\SI{20}{ppm}$ sulphur content. Melting front positions (\mbox{$f_\mathrm{L}$ = 1}) at $t = \SI{5}{\second}$ and values of $\mathrm{Pe^*}$ along the melting fronts, predicted using different permeability coefficients of \mbox{$C = 10^6\,\SI{}{\kilogram\per\second\per\meter\cubed}$}~(coloured lines), and~$10^{10} \,\SI{}{\kilogram\per\second\per\meter\cubed}$ (black lines): (\textbf{a})~$\Delta T_\mathrm{m} = \SI{200}{\kelvin}$; and (\textbf{b}) $\Delta T_\mathrm{m} = \SI{40}{\kelvin}$. Orange symbols show the experimental observation of the melting front position reported in~\cite{Pitscheneder_1996}.}
	\label{fig: weld_pool}
\end{figure}

\Cref{fig: weld_pool} shows the position of melting front ($f_\mathrm{L}$ = 1) at $t = \SI{5}{\second}$, as~well as the value of $\mathrm{Pe^*}$ along it when using different permeability coefficients and melting temperature ranges. The~laser-beam diameter is chosen here as the characteristic length scale $\mathscr{L}$ to calculate $\mathrm{Pe^*}$. Higher values of $\mathrm{Pe^*}$ are found for larger $\Delta T_\mathrm{m}$, predicting more sensitivity to the permeability coefficient as is indeed observed when comparing \cref{fig: weld_pool}a and \cref{fig: weld_pool}b. Due to a steep increase in the momentum sink term with liquid fraction for large permeability coefficients, all predictions are found to converge to an identical solution for very large $C$. However, a~very large value of the permeability can lead to numerical~instabilities.

\unskip

\section{Conclusions}
\label{sec: conclusion}

\noindent
A systematic study was performed to scrutinise the influence of the permeability coefficient on the numerical predictions of isothermal and non-isothermal phase-change simulations using the~enthalpy-porosity~method.

For isothermal phase-change problems, reducing the cell size diminishes the influence of the~permeability coefficient on the results, which become independent of the permeability coefficient for   fine enough meshes. A~grid independent solution of an isothermal phase-change problem is independent of the permeability coefficient. However, not every numerical result that is independent of the permeability coefficient is grid~independent.

Numerical predictions of non-isothermal phase-change problems are inherently dependent on the~permeability coefficient. The~sensitivity of the numerical predictions to the permeability coefficient increases with increased mushy zone thickness and increased fluid flow velocities perpendicular to the solid--liquid interface. A~method is proposed to predict and evaluate the influence of the~permeability coefficient on numerical predictions, and~verified for two-dimensional phase-change problems including laser spot melting. Large values of $\mathrm{Pe^*} \gg 1$ adjacent to the solid--liquid interface indicate a strong sensitivity and $\mathrm{Pe^*} \ll 1$ indicates insensitivity to the permeability coefficient $C$.

	\section*{Author Contributions}
	\label{sec:author_contributions}
	
	Conceptualisation, A.E., C.R.K. and I.M.R.; methodology, A.E.; software, A.E.; validation, A.E.; formal analysis, A.E.; investigation, A.E.; resources, A.E., C.R.K, and~I.M.R.; data curation, A.E.; writing---original draft preparation, A.E.; writing---review and editing, A.E., C.R.K.,  and~I.M.R; visualisation, A.E.; supervision, C.R.K. and  I.M.R.; project administration, A.E. and  I.M.R.;  and funding acquisition, I.M.R. 
	
	\section*{Acknowledgement}
	\label{sec:acknowledgement}
	
	This research was carried out under project number F31.7.13504 in the framework of the Partnership Program of the Materials innovation institute M2i (www.m2i.nl) and the Foundation for Fundamental Research on Matter (FOM) (www.fom.nl), which is part of the Netherlands Organisation for Scientific Research (www.nwo.nl). The authors would like to thank the industrial partner in this project “Allseas Engineering B.V.” for the financial support.

\bibliographystyle{elsarticle-num}
\bibliography{ref}

\end{document}